\documentclass[aps,reprint,article,prx]{revtex4-1}
\usepackage{amsmath}
\usepackage{dcolumn}
\usepackage{graphicx}
\usepackage{float}
\usepackage{color}
\usepackage{booktabs}
\usepackage{multirow}

\begin{document}

\title{Magnetic-field driven evolution of zero-energy mode on Bi islands deposited on Fe(Te,Se)}

\author{Kailun Chen, Chuanhao Wen, Zhiyong Hou, Huan Yang,$^*$ and Hai-Hu Wen$^\dag$}

\affiliation{National Laboratory of Solid State Microstructures and Department of Physics, Collaborative Innovation Center of Advanced Microstructures, Nanjing University, Nanjing 210093, China}

\begin{abstract}
We investigate the magnetic-field dependent evolution of the zero-bias conductance peaks (ZBCPs) on the nanoscale bismuth islands grown on the FeTe$_{0.55}$Se$_{0.45}$ substrate. The ZBCPs can be observed throughout the entire region on these islands, and their characteristics align with the signatures of Majorana zero modes. Remarkably, the evolution of ZBCPs on these islands exhibits anomalous behavior under varying magnetic fields: The magnitude of ZBCPs is first enhanced at weak fields lower than 2 T and then suppressed as the fields further increase. We attribute the non-monotonic evolution of the ZBCPs to the magnetic-field-enhanced topological edge states on these Bi islands. Our findings provide valuable insights into the probable origin of the Majorana zero modes in the Bi-island platform and the magnetic-field response of topological edge states.
\end{abstract}

\maketitle
\section{Introduction}

Majorana zero modes (MZMs) have attracted intensive interest because of their potential applications in fault-tolerant topological quantum computing \cite{SCZhangReview,AndoReview,NonAbelian}, and these modes can be realized in topological superconductors. One of the effective methods for achieving topological superconductivity is to induce superconductivity in the topologically insulating layer through the proximity effect from the adjacent superconductor \cite{LFuProximity}. For example, zero-energy modes were observed in vortex cores of the heterostructures composed by the topological insulator Bi$_2$Te$_3$ and the conventional superconductor NbSe$_2$ \cite{JiaReview,Jiaprl} or the iron-based superconductor Fe(Te,Se) \cite{ChenMYSA}. In addition, MZMs were also observed at terminals of vortex lines in superconductors with topologically non-trivial bands, such as some iron-based superconductors \cite{FeTeSeMajorana,FeTeSeHanaguri,HuReview,LiFeOHFeSe,CaKFe4As4,LiFeAs} or other materials \cite{WS2LiW,CVSWangZY}.

On the surface of a topologically non-trivial superconductor, topological edge states appear at some boundaries \cite{SCZhangReview,Hasan}, such as the twin boundary \cite{WangZYScience} or the step edge \cite{JiaoLUTe2}. In one-dimensional cases, topological edge states can exist in a semiconducting or spin-orbit-coupled nanowire, as well as in a ferromagnetic atom chain neighboring to a superconductor \cite{AliceaReview}, where they appear as MZMs. Experimental observations of these modes are realized at the ends of one-dimensional semiconducting nanowires \cite{SemiChain,Quantumdot} and magnetic atomic chains \cite{FeChain} grown on the surface of an $s$-wave superconductor. As for two-dimensional heterostructures, MZMs \cite{CrBr3} or Majorana edge modes \cite{Feisland,PbCoSi} are observed on ferromagnetic islands grown on $s$-wave superconductors. In addition, a robust zero-energy mode is observed in a trilayer heterostructure MnTe/Bi$_2$Te$_3$/Fe(Te,Se) \cite{ZhangT}.

Bismuth is a semimetal with strong spin-orbit coupling, and it is a good platform for investigating the topological superconductivity or the MZMs when it is adjacent to a superconductor. The Majorana edge states may exist at the boundary of the Bi layer with the former mentioned configuration. And the experimental evidence has been demonstrated at the edges of Bi bilayers \cite{Bibilayer} and Bi films decorated with magnetic iron clusters \cite{Edgechannel} grown on the superconducting substrate. In our previous research, robust zero-energy modes were observed on specific Bi islands deposited on the iron-based superconductor Fe(Te,Se) \cite{Biisland}. The zero-energy modes are likely caused by the interference of two counter-propagating topological edge states at the boundary of Bi islands. This kind of edge states in a topologically non-trivial system is protected by the time-reversal symmetry, and is impervious to impurity scattering in the absence of magnetic fields. Therefore, after applying varying magnetic fields, the evolution of the edge states is also an interesting issue that has been scarcely reported in experiments.

In this work, we examined the evolution of the ZBCP magnitude on some Bi islands grown on the FeTe$_{0.55}$Se$_{0.45}$ substrate when applying magnetic fields. Based on the statistic of the measuring areas, the ZBCPs exist only on some of the islands with a diameter of 4-8 nm and can be observed in the entire region of these islands. The characteristics of all the ZBCPs are also consistent with those of MZMs, indicating a topologically non-trivial origin of the ZBCPs. Notably, we observed an anomalous, yet general behavior of the ZBCPs upon varying magnetic fields: The intensity of the ZBCPs on Bi islands is first enhanced at weak fields lower than 2 T, and subsequently decreases as the fields further increase. The strengthening of the ZBCPs at weak fields may be attributed to the magnetic-field tuning on the edge states to the inner part of the island.

\section{Experimental methods}

The single crystals of FeTe$_{0.55}$Se$_{0.45}$ were synthesized by the self-flux method \cite{samplegrowth}. The crystals were annealed at $400^\circ$C for 20 h in an O$_{2}$ atmosphere to eliminate the interstitial Fe atoms and then quenched in the liquid nitrogen. The single crystal was cleaved in an ultrahigh vacuum with a pressure of about $1\times10^{-10}$ Torr before the growth of the Bi islands. High-purity Bi (99.999\%) powders were heated to $360^\circ$C in the effusion cell (CreaTec) and then evaporated to the cleaved surface of Fe(Te,Se) at room temperature by molecular beam epitaxy method. The nanoscale bismuth islands can be grown on the Fe(Te,Se) substrate. Afterwards the sample was transferred to the scanning tunneling microscopy/spectroscopy (STM/STS) head which was kept at a low temperature. The STM/STS measurements were carried out in a USM-1300 system (Unisoku Co. Ltd.) with an ultrahigh vacuum, low temperature, and high magnetic field. The tunneling spectra were measured by a lock-in technique with an amplitude of 0.2 mV and a frequency of 938 Hz. The tips in the measurements were made by tungsten using the electrochemically etching method. All measurements were taken at 0.4 K unless in some specified cases. The magnetic field was applied along the $c$-axis of Fe(Te,Se) substrate or equivalently perpendicular to the Bi islands.

\section{Results}

\subsection{Bi islands with and without ZBCPs}

\begin{figure}
\includegraphics[width=8.5cm]{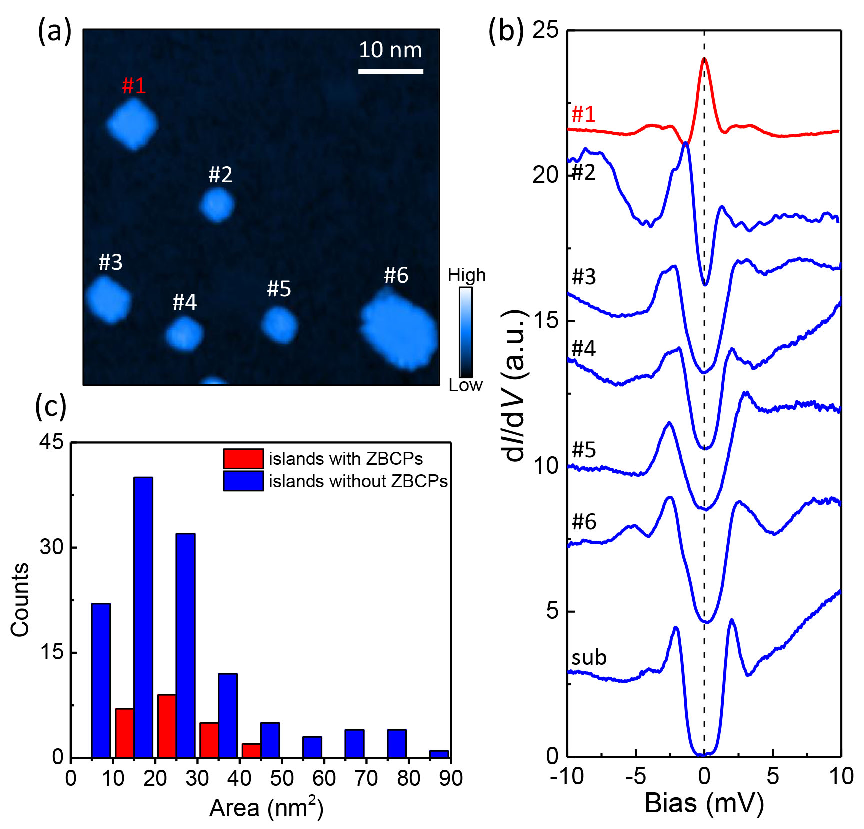}
\caption{(a) Topography of an area containing 6 Bi islands grown on the Fe(Te,Se) substrate (setpoint conditions: $V_\mathrm{set} = 1$ V, $I_\mathrm{set} = 20$ pA). (b) Typical tunneling spectra measured on the islands and the substrate nearby ($V_\mathrm{set} = 10$ mV, $I_\mathrm{set} = 200$ pA). The island \#1 is the only one with ZBCP. (c) Statistics on the number of islands with and without ZBCPs versus the area of all islands we have measured.}\label{fig1}
\end{figure}

Figure~\ref{fig1}(a) shows a typical topography of the Bi islands grown on the Fe(Te,Se) substrate. The islands are randomly distributed on the flat surface of the substrate, with the dimensions of several nanometers. The shapes of these islands are also irregular, and there are even some wrinkles near the boundary indicating the lattice distortion there. Although the islands have different sizes, the height of them is all about 7 \AA\, which is consistent with the thickness of Bi(110) monolayer islands \cite{Biheight}. These features of the Bi islands are consistent with those in our previous work \cite{Biisland}.

Figure~\ref{fig1}(b) shows typical tunneling spectra measured on the six islands in the field of view of Fig.~\ref{fig1}(a). We also present the typical tunneling spectrum measured on the Fe(Te,Se) substrate in Fig.~\ref{fig1}(b), and it shows a fully gapped feature. The superconducting gap of the Fe(Te,Se) substrate varies from 1.1 to 2.1 meV determined by calculating the energy difference between coherence peaks \cite{ChenMYCdGM}. Tunneling spectra measured on Bi islands \#2-\#6 are similar to the one measured on the Fe(Te,Se) substrate. In contrast, the tunneling spectrum measured on island \#1 is different, i.e., a ZBCP appears in the tunneling spectrum measured on this island. It should be noted that the ZBCP can be observed in the spectra measured on the whole island \cite{Biisland}. We have investigated 146 islands with monolayer thickness, and only 23 of them exhibit ZBCPs. The probability of finding a Bi monolayer island with the ZBCPs is about 16\%. We also note that there are some bilayer Bi islands, but none of them hold the ZBCPs. The number statistics are also carried out in the region of monolayer islands, and the result is shown in Fig.~\ref{fig1}(c). One can see that the areas of islands with ZBCPs are mainly distributed from 10 nm$^2$ to 50 nm$^2$, corresponding to the diameter of about 4-8 nm. When the areas of Bi islands exceed 50 nm$^2$, no ZBCPs has been observed in these islands.

\begin{figure}
\includegraphics[width=8.5cm]{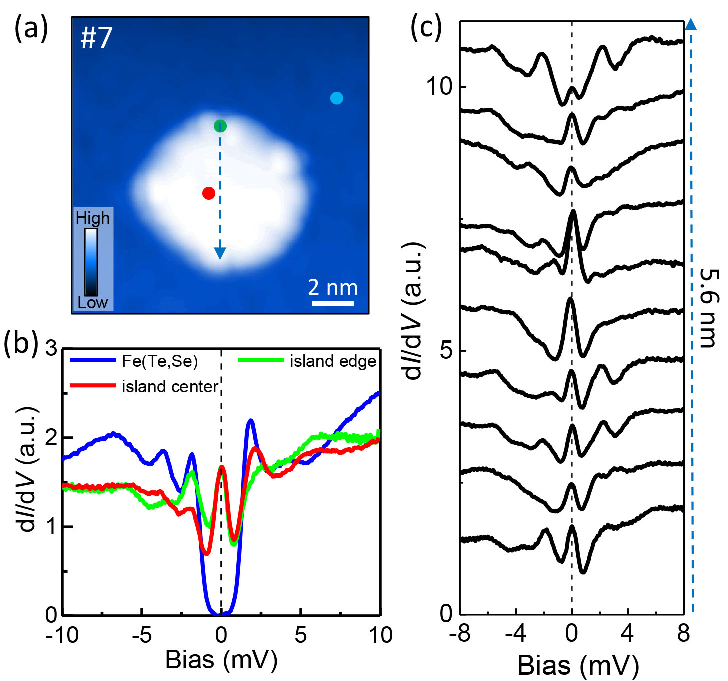}
\caption{(a) Topography of a Bi island (\#7) ($V_\mathrm{set} = 1$ V, $I_\mathrm{set} = 20$ pA). (b) Typical tunneling spectra measured at marked positions shown in (a) ($V_\mathrm{set} = 10$ mV, $I_\mathrm{set} = 200$ pA). (c) Line-profile of tunneling spectra taken along the dashed arrow in panel (a) ($V_\mathrm{set} = 10$ mV, $I_\mathrm{set} = 200$ pA).}\label{fig2}
\end{figure}

Figure~\ref{fig2}(a) shows the topography of a nanoscale monolayer Bi island numbered as \#7. The ZBCPs can be observed in the tunneling spectra measured on the island, such as two spectra measured at the edge and the center of the island shown in Fig.~\ref{fig2}(b). Besides, the energies of the coherence peaks of these two spectra are similar to those obtained from the spectra measured on the Fe(Te,Se) substrate, which is a demonstration of the proximity-induced superconductivity on the Bi island. It should be noted that ZBCPs can be observed in the spectra taken all over the island, and one can obtain the conclusion from a set of tunneling spectra measured across the island as shown in Fig~\ref{fig2}(c). Obvious in-gap peaks can be seen in all the spectra in this panel, and the peak positions are fixed near zero energy. These observations are similar to those in our previous work \cite{Biisland}.

\subsection{Magnetic-field dependent evolution of ZBCPs on Bi islands}

\begin{figure}
\includegraphics[width=8.5cm]{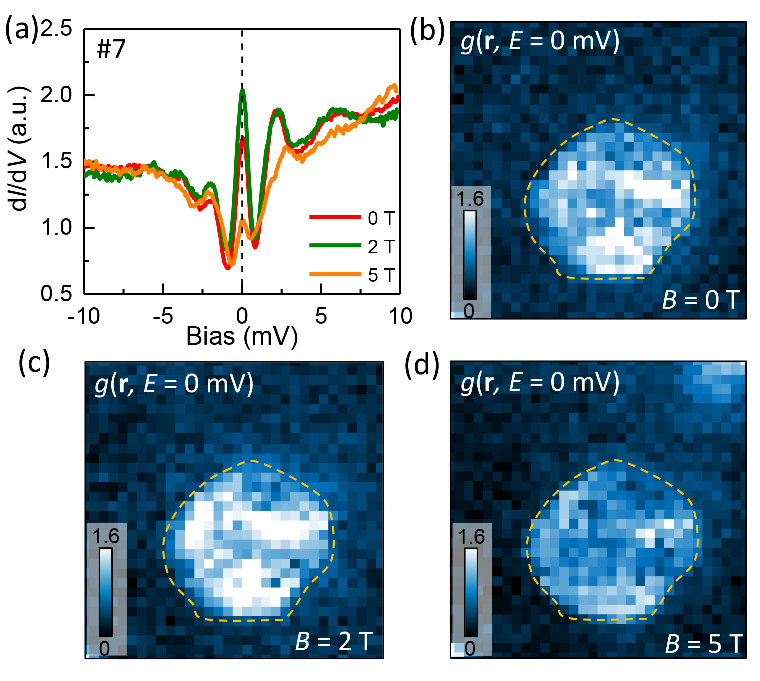}
\caption{(a) Tunneling spectra measured at different magnetic fields and at the location of red dot in Fig.~\ref{fig2}(a) ($V_\mathrm{set} = 10$ mV, $I_\mathrm{set} = 200$ pA). (b)-(d) Zero-energy d$I$/d$V$ mappings recorded in the same region in Fig.~\ref{fig2}(a), and they are measured at different fields ($V_\mathrm{set} = 40$ mV, $I_\mathrm{set} = 200$ pA). The edge of the island is marked out by dashed lines. The color bars are the same in these mappings.}\label{fig3}
\end{figure}

The ZBCPs on some Bi islands are weakened by increase of the temperature \cite{Biisland}, which can be understood as the temperature suppression to the superconductivity in the Fe(Te,Se) substrate. Since the upper critical field of Fe(Te,Se) is extremely high \cite{FeSeTeHc2}, it is interesting to investigate the field-dependent evolution of the ZBCPs on the Bi islands. Figure~\ref{fig3}(a) shows tunneling spectra measured at different fields of 0, 2, and 5 T at the center of the Bi island \#7. Surprisingly, the increment of the magnetic field does not suppress the ZBCP monotonically, and conversely the peak magnitude increases at 2 T compared to that obtained at 0 T. At a higher field of 5 T, the magnitude of the ZBCP is significantly suppressed, but the ZBCP does not show any splitting or broadening features.

The differential conductance mapping is a useful method to get the information about the spatial distribution of density of states (DOS) \cite{STMReview1,STMReview2}. Figures~\ref{fig3}(b)-\ref{fig3}(d) show the recorded spatial distributions of zero-bias differential conductance of the Bi island \#7 in the same area but under different fields. These three mappings are presented in the same color scale, thus the brightness directly corresponds to the zero-energy DOS. The zero-bias differential conductance is almost zero on the Fe(Te,Se) substrate, reflecting the fully gapped feature. The value is finite on the whole island at 0 T, which corresponds to the robust zero mode on the island. Some weak ZBCP magnitude may be due to the surface-lattice distortion of the Bi islands. At the magnetic field of 2 T [Fig.~\ref{fig3}(c)], the inner part of the island becomes notably brighter, suggesting an increment of the ZBCP magnitude. However, at 5 T [Fig.~\ref{fig3}(d)], the zero-energy differential conductance becomes much weaker. These observations are consistent with the ZBCP evolution in the tunneling spectra at different fields shown in Fig.~\ref{fig3}(a).

\begin{figure}
\includegraphics[width=8cm]{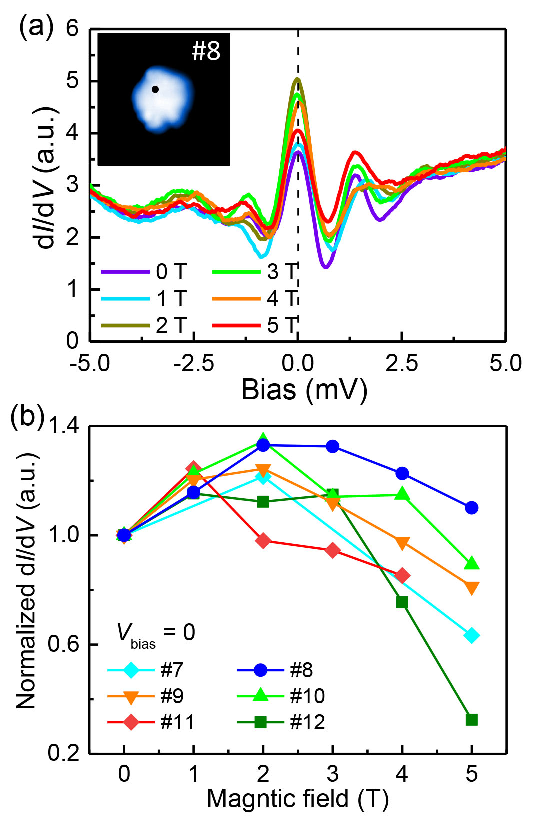}
\caption{(a) Tunneling spectra measured on another island \#8 at different magnetic fields ($V_\mathrm{set} = 10$ mV, $I_\mathrm{set} = 200$ pA). The topography of the island is shown in the inset, and the spectra are measured at the marked position in the island. (b) Magnetic field-dependent evolution of the ZBCP intensity on several islands with ZBCPs. The values of the zero-energy differential conductance at different fields are normalized by the value at zero field.
} \label{fig4}
\end{figure}

A control experiment is carried out on the Bi island \#8, and Fig.~\ref{fig4}(a) shows the tunneling spectra measured at the same position but under different magnetic fields. The magnitude of the ZBCP increases when the magnetic field increases from 0 to 2 T, and then the value decreases as the field further increases. Similar experiments are also carried on other four islands \#9-\#12 with ZBCPs, and the field-dependent zero-bias differential conductance is shown in Fig.~\ref{fig4}(b). Here it should be mentioned that these data are recorded in the Bi islands away from vortex cores in the Fe(Te,Se) layer, otherwise the tunneling spectra behave very differently from the spectra measured at other neighbored fields. The curves in Fig.~\ref{fig4}(b) share similar features with varying magnetic fields: the intensity of the ZBCP increases with increasing field and reaches its maximum at about 2 T, and afterward the ZBCP magnitude decreases rapidly with the increase of the field. Therefore, the non-monotonic field-dependent evolution of the ZBCP magnitude is a common property on these Bi islands grown on Fe(Te,Se).

\section{Discussion}

As presented above, we have investigated the magnetic-field dependence of the ZBCP magnitude in some monolayer Bi islands grown on the superconducting FeTe$_{0.55}$Se$_{0.45}$ substrate. On these specific islands with the size of 4-8 nm, the zero-energy peak is the common feature of the spectra measured throughout the entire island. In the process of applying magnetic fields, the ZBCPs are fixed at zero energy and do not split. Additionally, the width of the ZBCPs does not broaden. These features exclude the trivial origin of the ZBCPs caused by Yu-Shiba-Rusinov states or the Andreev bound states \cite{ChenXY}, and are consistent with the characteristic of MZMs as reported previously \cite{Jiaprl,IFI,ChenXY,Adatom}. Thus, the ZBCPs on the Bi islands probably have a topologically non-trivial origin. As discussed before \cite{ChenXY}, the ZBCPs may be due to the magnetic moment just below the particular Bi island. The magnetic moment can be induced by an interstitial iron atom, and it leads to the time-reversal symmetry breaking \cite{AliceaReview}. However, in the present work, the ZBCP magnitude increases with the increase of magnetic fields and reaches its maximum at about 2 T; afterwards, the magnitude decreases with further increase of the fields. This is very different from the situation of excess iron impurities, and the ZBCP magnitude is robust at a field as high as 8 T \cite{IFI}. In addition, the effective range by the excess iron atom is very limited with a radius of about 1 nm \cite{IFI} which is much smaller than the size of the Bi island. Therefore, the excess iron atom as the origin of the ZBCP may be excluded, and the ZBCP is likely caused by the topological superconductivity induced on the Bi island with a strong spin-orbital coupling.

From our previous study \cite{ChenXY}, the zero-energy states observed on these Bi islands may be caused by two counter-propagating topological edge states \cite{TSC3D} emerging at the edge of the islands. The edge states usually behave as the spatial oscillation as a form of the Bessel function, and the real-space period approximately equals to $\pi/k_\mathrm{F}$ \cite{HaoN1}. Since the Fermi vector $k_\mathrm{F}$ is very small in Bi, the real-space period can be the scale of several nanometers. When the size of the island is suitable, the pair of the edge states may form an interfered resonant state which behaves as the ZBCP. Based on this picture, the applied magnetic fields can tune the real-space period of the oscillation as well as the decaying parameter of the edge state to the inner part of the island \cite{HaoN1}, which may help to increase the ZBCP magnitude at fields lower than 2 T. In comparison, the ZBCP magnitude decreases monotonically with increase of the magnetic field in the situation of the iron impurity \cite{IFI,HaoN2}, the magnetic monolayer film in the trilayer heterostructure \cite{HaoN1} and the vortex core in Fe(Te,Se) \cite{ChenXY}.

Another possibility of the non-monotonic evolution of the ZBCP magnitude is from the vortex core in the Fe(Te,Se) substrate. Although our tunneling spectra are recorded when the Bi island is away from the vortex core in Fe(Te,Se), there are also some vortex cores nearby. One of them are shown in the upper right corner of Fig.~\ref{fig3}(d). The superconducting current surrounding the vortex cores may pass through the Fe(Te,Se) underneath, which may probably enhance the edge states on the Bi island and induce an increment of the ZBCP magnitude. At higher magnetic fields, the proximity-induced superconductivity is strongly suppressed by the fields, and the ZBCP magnitude decreases. Clearly, further theoretical consideration is highly desired to understand the non-monotonic relationship between the ZBCP magnitude and the magnetic field quantitatively, as well as the reason why ZBCPs only appear on specific Bi islands with particular sizes/shapes.

\section{Conclusion}

In summary, we have observed the zero energy modes on certain Bi islands with the diameter of 4-8 nm deposited on the FeTe$_{0.55}$Se$_{0.45}$ substrate. These zero energy modes may be  MZMs and can be found throughout the entire region on these islands. Further measurements on these islands under varying magnetic fields reveal an unusual behavior of the MZM magnitude. Specifically, the magnitude is initially enhanced at weak fields lower than 2 T, but then becomes suppressed as the field strength increases. Concerning the probable origin of the zero-energy modes we discussed above, the anomalous evolution is probably caused by the magnetic-field tuning on the edge states or influenced by the superconducting current surrounding the vortex cores on the Fe(Te,Se) substrate. Our findings will stimulate theoretical efforts for understanding the mechanism of the bismuth topological systems and shed new light on the topological property of Majorana zero modes under varying magnetic fields.

\begin{acknowledgments}
We appreciate very useful discussions with J{\"o}rg Schmalian, Ning Hao, and Haijun Zhang. The work was supported by the National Natural Science Foundation of China (Grants No. 11974171, No. 12061131001, and No. 11927809) and the National Key R\&D Program of China (Grant No. 2022YFA1403201).
\end{acknowledgments}

$^*$ huanyang@nju.edu.cn,
$^\dag$ hhwen@nju.edu.cn

\end{document}